\documentclass{article}
\usepackage{emulateapj}
\usepackage{apjfonts}
\usepackage{timesfonts}
\usepackage{psfig}

\begin{document}

\title{On The Spectral Energy Dependence of Gamma-Ray
Burst Variability}   

\author{Nicole M. Lloyd-Ronning\altaffilmark{1} and Enrico Ramirez-Ruiz\altaffilmark{2}}
\affil{ $^{1}$ Canadian Institute of Theoretical Astrophysics, 60 St. George
Street, Toronto, M5S 3H8, Canada; lloyd@cita.utoronto.ca}
\affil{ $^{2}$ Institute of Astronomy, Madingley Road, Cambridge,
CB3 0HA, England; enrico@ast.cam.ac.uk}

\begin{abstract}
The variable activity of a $\gamma$-ray burst (GRB) source is thought to
be correlated with its absolute peak luminosity - a relation that, if
confirmed, can be used to derive an independent estimate of the
redshift of a GRB. We find that bursts with highly variable light
curves have greater $\nu F_{\nu}$ spectral peak energies, when we
transform these energies to the cosmological rest 
frame using the redshift estimates derived either from optical
spectral features or from the luminosity-variability distance indicator
itself. This positive correlation between peak
energy and variability spans $\approx$ 2 orders of
magnitude  and appears to accommodate GRB 980425, 
lending credibility to the association of this burst with SN 1998bw.
The existence of such a correlation not only provides an
interesting clue to the nature of this luminosity indicator but
potentially reinforces the validity of the 
redshift estimates derived from this method. It also implies that
the rest frame GRB peak energy is correlated with the intrinsic
luminosity of the burst, as has been suggested in the past as an
explanation of the observed hardness-intensity correlation in
GRBs. 

\end{abstract}

\section{Introduction}
Until a few years ago, gamma-ray bursts (GRBs) were known only as brief,
intense flashes of high-energy radiation, with no observable
traces at other wavelengths. The study of fading afterglows has
enabled the measurement of redshift distances and the identification of
host galaxies, establishing that GRBs are extremely
luminous events detectable to much larger distances than
quasars or galaxies.  
Consequently, GRBs can provide novel information about early epochs in the
history of the universe.

Present  distance estimates  - which rely on optical
line features in the afterglow spectrum or emission lines
in the spectrum of the host galaxy - are relatively rare. 
There are now $\sim 20$ GRBs with optical spectroscopic redshifts (all
in the range 0.43 $\le \,z \, \le$  4.5).\footnote{See Jochen Greiner's
page at {\tt http://www.aip.de/~jcg/grb.html} for an excellent compilation of 
information on GRBs with redshifts.} Hence, until recently it
appeared that using GRBs to map 
the high-$z$ universe would have to wait for dedicated localization
and follow-up programs like {\it Swift}\footnote{\tt
http://swift.gsfc.nasa.gov} and {\it NGST}\footnote{\tt
http://ngst.gsfc.nasa.gov}. However, the discovery of two
recent correlations between the degree of variability of the
$\gamma$-ray light curve and the GRB luminosity (Ramirez-Ruiz \&
Fenimore 1999; Fenimore \& Ramirez-Ruiz 2001), and between the
differential time lags for the arrival of burst pulses at 
different energies and the luminosity (Norris, Marani \& Bonnell 2000) offer
the possibility of deriving independent estimates of the 
redshift of  a GRB. Interestingly,  in a large sample of BATSE bursts,
the time lags for the arrival of burst pulses at  different energies
and the degree of variability appear to be strongly related
(Schaefer, Ming \& Band 2001), lending credence to each correlation.
While these correlations are still tentative, they seem to be a natural
consequence of the variation in energy per unit solid angle  (or bulk Lorentz
factor)
 of the emitting region (Salmonson 2000; Ioka \&
Nakamura 2001; Kobayashi, Ryde \& MacFadyen 2001; Plaga 2001; Ramirez-Ruiz \& Lloyd-Ronning, 2002).
 That is, an increase in energy per unit solid angle 
 through, for example, an increase
in the relativistic source expansion velocities can lead to more luminous
bursts as well as shorter observed timesales in accordance with the observed
correlations (see Ramirez-Ruiz \& Lloyd-Ronning 2002). \\
 
Here we show that - when transformed to the cosmological
rest frame using redshifts derived either from spectroscopic observations or
from the luminosity-variability ($L-V$) relation - bursts with highly variable
light curves have greater typical peak energies.
The paper is organized as follows: In \S 2, we present the intrinsic peak energy-variability
correlation. In \S 3, we discuss possible observational selection effects that
may affect this correlation and show that - even when assuming the most conservative,
severe data truncation - the correlation still holds to high statistical significance.
In \S 4, we briefly discuss how a wide variety of burst
phenomenology may be attributable to the existence of this
correlation (in conjunction with the $L-V$ relation). We also suggest that
this result not only provides useful insight into the physics of the GRB
mechanism, but also may support the validity of the $L-V$
relation as a reliable luminosity indicator. Finally, we present our conclusions in \S 5.
Throughout our analysis, we assume $H_0 = 65\,\, {\rm km} \, {\rm s}^{-1} \, {\rm
Mpc}^{-1}$, a matter density $\Omega_{\rm matter}=0.3$, and a vacuum
energy density $\Omega_{\Lambda}=0.7$.         

\section{GRB Spectra and Variability}
GRB temporal profiles are so enormously varied and complicated that, at first
sight, their behavior obeys no simple rule. Many bursts have a highly
variable temporal profile with a timescale of variability that is
significantly shorter than the overall duration. Several studies have
suggested the possibility of relating properties of the time structure
with the burst luminosity (Stern, Poutanen \& Svennson 1997;
Beloborodov et al. 2000; Norris et al. 2000; Fenimore \& Ramirez-Ruiz 
2001; Reichart et al. 2001). In particular,  Fenimore \& Ramirez-Ruiz 
(2001) explored the possibility of using the ``spikiness'' of the time
structure, combined with the observed flux, to obtain distances, much as the Cepheid relationship
gives distances from the pulsation period. 
Several hundred long and bright bursts were amenable
for their analysis, producing a large sample of events with derived
redshifts  and luminosities. Besides using this large sample to
understand both the  intrinsic GRB luminosity function and GRB
formation rate (e.g. Fenimore \& Ramirez-Ruiz  2001;  Lloyd-Ronning, Fryer \&
Ramirez-Ruiz 2002), these estimates offer the possibility of studying the
physical nature of the luminosity indicator itself. 
To that effect, we investigate the dependence of the burst spectra on
variability. From 
the set of 220 bright, long, BATSE bursts that 
Fenimore \& Ramirez-Ruiz  (2001) analyzed, we use all 159 that have time
resolved fits from 16 channel spectral data
(R.S. Mallozzi\footnote{Deceased.}, private communication). 
The observed spectra are phenomenologically well
characterized by the ``Band'' function (Band et al. 1993), defined by
a low-energy spectral 
index, $\alpha$, a high-energy spectral index, $\beta$, and the peak
of the $\nu F_\nu$ distribution, $E_p$\footnote{The parameter $E_p$
corresponds to the peak of the spectrum in  $\nu F_\nu$ only if
$\beta$ is less than -2. Otherwise, the spectral peak is given by the
lower boundary in energy of the high-energy power-law component
characterized by $\beta$ (Preece et al. 2000).} (at which the source
is observed to emit the bulk of its luminosity).  We use the spectral parameters
from the time when each burst's photon flux is maximum (i.e.the burst's ``peak''), but
find qualitatively similar results if time averaged spectra are used.  Table 1 lists
the data we have used in our analysis.

Figure 1 shows the GRB peak energy in the cosmological rest frame,
$E_{p'}=E_p(1+z)$, versus the observed variability, $V$. The filled
circles are BATSE bursts with secure redshifts, high-resolution
light curves and resolved  spectral fits (Fenimore \& Ramirez-Ruiz
2001; Jimenez, Band \& Piran 2001), while the open circles are bursts
with redshift estimates derived from the $L-V$
indicator itself.  We find
a significant ($\gtrsim 5 \sigma$) positive correlation between
$E_{p'}$ and $V$ that extends for about 
two orders of magnitude; this correlation - taken at face value - can
be parameterized as $E_{p'} \propto V^{\approx 0.8 \pm 0.2}$. However,
we caution that 
the quantitative estimate of this correlation can be  
affected by selection effects (indicated by the shaded regions and solid line
in Figure 1).  We discuss this in detail in \S 3 below. Of the bursts in our
sample, GRB 980425 is unique because of its possible association with
SN 1998bw (Galama et al. 1998). The fact that it is consistent with
the observed trend (see Figure 1) suggests that this event and the cosmological
bursts may share a common physical origin. This speculation is made
more intriguing by a recent discovery that, at least in some bursts,
a supernova may be involved (Bloom et al. 1999; Reichart 1999; Lazzati
et al. 2001) which may have contributed
to an otherwise unexplained bump and reddening in the optical
light curve (but see Esin \& Blandford 2000 and
Ramirez-Ruiz et al. 2001 for  alternative explanations).

It is also important to note that each burst's redshift used to transform the
peak energy into the cosmological rest frame is derived from the luminosity
indicator itself (apart from the bursts with secure $z$ that are used to
calibrate the correlation).  As more GRBs with independent spectroscopic
redshifts are obtained, the existence of this (and the $L-V$)
correlation will be more definitively tested. So far, however, 
those bursts with secure redshift
estimates seem to fall well along this trend as seen in Figure 1. If
we fit a power law to the correlation for just the 7 (or 8) bursts with
measured redshifts, we
find $E_{p'} \propto V^{0.75 \pm 0.3}$   
 and $E_{p'} \propto V^{0.45 \pm 0.15}$, including and excluding GRB 980425 respectively.  The
existence of the $E_{p'}-V$ correlation in our sample of 159
GRBs, therefore, may provide
some confidence in the validity of the redshifts derived from the
$L-V$ luminosity indicator.\\

 As an illustration, in Figure 2 we show a histogram of the rest frame peak
energy $E_{p'}$, for those 159 bursts which have
spectral fits and redshifts from the $L-V$ relation.  The superposed dotted
histogram shows the observed peak energy for reference.
Note that the distribution of $E_{p'}$ peaks at about $1 {\rm MeV} \sim 2
m_{e}c^{2}$. If one believes the redshifts from this luminosity indicator,
then we can use this {\em intrinsic} distribution of GRB spectral peak
energies to gain insight into the relevant particle acceleration
processes and emission mechanisms present in GRBs. These possibilities
and their consequences for the predicted prompt and afterglow
emissions are investigated in Ramirez-Ruiz \& Lloyd-Ronning (2002).
One should keep in mind that the observed $E_{p}$ distribution
in Figure 2 is for a limited sample of bright bursts observed by
BATSE and in particular does not include the 
increasing number of bursts with values of $E_{p}$ that fall 
below the BATSE threshold, which possibly account for up to 1/3 of
all GRBs (e.g. Kippen et al. 2001, Heise et al. 2001).
  As of yet, these so-called ``X-ray Flashes'' (XRFs)
have no quantitative variability measurements (due to their very low
fluxes, at least in the BATSE data) although it has been qualitatively claimed
that they exhibit rapid variation in their time profiles, representative of the 
``typical'' BATSE population (Kippen et al. 2001).  
  It will be interesting to see if these bursts follow the
trend exhibited in Figure 1. We note that all analysis to
date of these bursts indicates that they tend to at least marginally exhibit the
same trends as the bulk of the BATSE bursts (see, e.g., Kippen et al. 2001).
However, once variability measurements of these bursts are made, this correlation
should be re-examined, including this sample.
\\ 

\section{The Role of Selection Effects}
We realize some of
the limitations that are intrinsic to our procedure. 
Before drawing any conclusions from
the $E_{p'}-V$ correlation, it is essential to understand the role
selection effects play in determining this trend, as well
as the  uncertainties imposed by the scatter in the
$L-V$ relation.  We discuss each in turn below. 

\subsection{Truncation due to Flux Limit}
The  $L-V$
sample is a selected sample of bursts above a
flux threshold of $1.5 \rm ph/cm^{2}/s$, and a duration threshold $>
20$ s. This  latter selection criterion is unlikely to play an
important role in our analysis since $L$ and 
$E_{p'}$ are relatively independent of duration. On the other hand, 
the  flux threshold has the effect of causing a very strong truncation in the
$L-z$ plane, in the sense that low luminosity bursts at
high redshift are  not ``observed'' (see  Figure 2 in
Lloyd-Ronning et al. 2001). 
Because $L$ and $V$ are correlated, this translates to
a truncation in the $V$-$z$ plane.  In
other words, the flux selection criterion excludes bursts with
low variability at high redshift. We would like to understand where these
observationally ``missing'' bursts might fall in the $E_{p'}-V$ plane.
 Now, because our {\em observed} $E_{p}$
distribution is relatively narrow and uncorrelated with redshift, when
we transform into the cosmological rest frame $E_{p'}=E_{p}(1+z)$, 
we find (on average) higher $E_{p'}$ values at higher
redshifts.  Therefore there is a selection against low variability,
high peak energy bursts (the upper left of the $E_{p'}-V$ plane).
We investigate the role - if any - this selection effect plays in producing the observed
correlation.  

  A quantitative formulation of the truncation in the $E_{p'}-V$ plane
from the luminosity-redshift selection effect is not straightforward
(primarily because  the scatter in the observed $L-V$ relation does
not allow for an unambiguous 
change of variables between $L$ and $V$).  However, we can make an
estimate of this truncation if we assume a perfect correlation between
$L$ and $V$, $L \propto V^{3.3}$ (this is the best-fit relation
derived by Fenimore \& Ramirez-Ruiz 2001).  From the flux threshold selection
criterion, a truncation is produced in the luminosity-redshift plane: 
$L_{lim} \propto f_{p,lim} d_{z}^{2}$, where
$d_z = (1+z)\int_{0}^{z} dz/\sqrt{\Omega_{\Lambda} +\Omega_{m}(1+z)^{3}}$, and
$f_{p,lim}$ is the peak flux limit that Fenimore and Ramirez-Ruiz chose - $1.5$
ph/cm$^{2}$/s.  We can then estimate the limiting variability as a function
of $E_{p'}$ by changing variables from $L$ to $V$ (using $L \propto V^{3.3}$)
and the fact that $E_{p'}/E_{p} = (1+z)$.  We find
 $V_{\rm lim}(E_{p'}) \propto [E_{p'}
\int_{1}^{E_{p'}/E_{p}}dx \ (\Omega_{\Lambda} + \Omega_{\rm m}
x^{3})^{-1/2}]^{2/3.3}$. This estimate of the truncation is in fact
quite shallow as shown by the solid line in Figure 2.  I has
no effect on the quantitative results
reported in section 2 (which did not account for selection effects).
This point is further illustrated in Figure 3, which shows the intrinsic
peak energy $E_{p'}$ vs. V for increasing flux bins 
(the squares, triangles, pentagons, hexagons, and
horizontal lines are
for $f_{p} = 1.5-2.2$ ph/cm$^{2}$/s,  $f_{p} = 2.2-2.7$ ph/cm$^{2}$/s,
 $f_{p}=2.7-4.0$ ph/cm$^{2}$/s, 
$f_{p} = 4.0-7.0$ ph/cm$^{2}$/s, and $f_{p} > 7.0$ ph/cm$^{2}$/s, respectively).
If the flux limit caused a severe
bias against observing bursts in the upper left hand corner of the $E_{p'}-V$ plane, 
 we expect to see these points migrate toward
 the right lower corner of the $E_{p'}-V$ plot, with increasing flux bin. 
This is clearly not the case, and in fact the points are scattered throughout
one another in the plot.

In the analysis that follows, we in fact take the most conservative
approach by imposing a {\em steeper}, more severe truncation than what
the analytical estimate in the preceeding paragraph predicts. 
We feel this more conservative approach not
only places our results on much firmer statistical ground, but  also
implicitly accounts for the effects of scatter in the $E_{p'}-V$ plane. 

\subsection{Accounting for the Selection Effects}
As mentioned above, it is important to quantify the role of selection
effects in the data.  Fortunately, there are firmly established
non-parametric statistical methods that have been developed to deal
with precisely such selection effects (e.g. Lynden-Bell 1971;
Efron \& Petrosian 1992). 
These techniques use a well-defined truncation criterion (and the assumption
that the observed sample is the most likely to {\em be} observed) to
estimate the correlation between (and underlying parent distributions
of) the relevant variables.  For each data point indexed by $i$, an
``eligible set'' is defined based on those points that fall within the
observational limits of the $i$'th data point at hand.  This amounts to making
a truncation parallel to the axes for each data point; a weight is then
assigned to each point given the number of points in its eligible
set. For example, an
eligible set for the $i$'th data point in our sample consists of all
bursts indexed by $j$, where $V_{lim, j} < V_{i}$ and $V_{j} > V_{i}$. Correlations
are then computed via non-parametric rank statistics (such as a Kendell's
$\tau$ test), where rank comparisons are made only among   
those within the eligible sets. Further details of these techniques
can be found in Efron and Petrosian (1992) and in the appendix of
Lloyd, Petrosian, \& Mallozzi (2000). 

As mentioned in \S 3.1, to estimate how the peak flux selection
criterion affects our results, we take the most conservative view that
{\em all} of the lack of bursts in the upper left corner is the result
of an observational selection effect. We in fact employed two
estimates of this truncation for our data - shown by the two shaded 
regions in Figure 1.  The first (dark shaded region) fits as closely
to the data as possible without eliminating any points; the latter,
more severe truncation (light shaded region) artificially eliminates
``high scatter'' points, to allow the truncation to fit tightly to the
majority of the data. For the first truncation,
we find a significant $ > 5 \sigma$ correlation
between $E_{p'}$ and $V$. The functional form of this correlation  
(accounting for the truncation of course) can be expressed as $E_{p'}
\propto V^{0.8 \pm 0.15}$. For the second truncation, we again find a
$> 5 \sigma$ correlation between $E_{p'}$ and $V$, which can be parameterized
as $E_{p'} \propto V^{0.7 \pm 0.15}$.   The existence of the
correlation, therefore, is likely to primarily be due to
the lack of bursts with high variability and low $E_{p'}$ (i.e. the
lower right hand corner of Figure 1) which we believe is real and is
not a result of any selection effect.  We again emphasize that the flux limit
most likely imposes much less severe truncation than we have assumed here (see
solid line in Figure 1).

\subsection{Consequences of Uncertainties in the $L-V$ Relation}
We have also computed the correlation between $E_{p'}$ and $V$ given
the luminosities and redshifts derived from the upper and lower limits
to the fitted $L-V$ relation. This correlation is parameterized as $L
\propto V^{\beta}$ with $\beta = 2.2, \ 3.3$, and $5.8$ for the lower
limit, best-fit, and upper limit respectively (see Fenimore \&
Ramirez-Ruiz 2001). As for the best fit relation (see \S 3.2 and Figure
1), we chose two truncations for each data set -- the first fitting as
closely as possible to the data without eliminating any data points
(truncation 1), and the second eliminating ``high scatter'' points
(truncation 2). In all cases, we find a $> 4 \sigma$ correlation
between $E_{p'}$ and $V$.  For  $\beta = 2.2$, the correlation can be
parameterized by $E_{p'} \propto V^{0.5 \pm 0.15}$ (truncation 1) and
$E_{p'} \propto V^{0.4 \pm 0.15}$ (truncation 2). For $\beta = 5.8$, the 
parameterization is $E_{p'} \propto V^{1.15 \pm 0.15}$ (truncation 1)
and $E_{p'} \propto V^{1.05 \pm 0.15}$ (truncation 2). All of these results are
summarized in Table 2.\\

It is also important to notice that we are relying on redshifts 
that are obtained from the $L-V$ relation, which we have
extrapolated to higher variabilities than those from
which the correlation was derived. We have therefore computed the
correlation between $E_{p'}$ and $V$ for only those bursts in our
sample which fall in the variability range corresponding to the
observed bursts that were used to calibrate the $L-V$ correlation.
Again, accounting for selection effects, we still find a highly
signficant correlation ($> 5 \sigma$) between $E_{p'}$ and $V$ for
bursts in this limited range (both including 
and excluding the variability of GRB 980425 in defining the lower
limit of our range), with $E_{p'} \propto V^{0.9 \pm 0.2}$.
 In addition, we have computed the correlation between
$E_{p'}$ and $V$ for only those bursts in our sample which fall in the
redshift range corresponding to the observed bursts that were
used to calibrate the $L-V$ correlation, $0.8 < z < 3.4$ (note that we
exclude GRB 980425; including this burst in our redshift range will
increase the significance of our results).  Once again, we find a $\ga
4 \sigma$ correlation  between $E_{p'}$ and $V$ for the $\approx 80$
bursts that made this redshift cut, with $E_{p'} \propto V^{0.75 \pm 0.2}$. 
 We conclude from this analysis that
if the $L-V$ relation holds true (even over a limited range),
then highly variable bursts have greater intrinsic peak energies.

\section{On the relationship between temporal and spectral structure
in GRBs} 

The existence of the $E_{p'}-V$ correlation also implies the (not
surprising) fact that the GRB luminosity and intrinsic peak energy are
correlated, as directly follows from the $L-V$ and $E_{p'}-V$
relations.  We indeed find such a correlation in our data at high ($> 5 \sigma$)
 statistical
significance.
This result is qualitatively consistent with the findings of Lloyd,
Petrosian \& Mallozzi (2000), who 
suggested that the observed hardness-intensity correlation in GRBs is
an intrinsic (and not cosmological) effect - namely, a correlation
between GRB rest frame peak energy and luminosity.

 Moreover,  several
mutually reinforcing trends have been found in the past that support
the validity of our results. One key trend
involves the tendency for pulses or ``peaks'' in GRB time histories to
be narrower at higher energies. This was first noted by
Fishman et al. (1992), Link,
Epstein \& Priedhorsky (1993) and quantitatively explored by
Fenimore et al. (1995).  The latter showed that the average pulse
width has a power-law dependence on energy with an
index of about -0.4.  A visual inspection
of the pulses fitted to GRBs by Norris et al. (1996) shows that the
low-amplitude pulses (within a single burst) tend to be
wider. Finally, Ramirez-Ruiz \& Fenimore (2000) found a quantitative
relationship between pulse amplitude and pulse width: the smaller
amplitude peaks tend to be  wider with the pulse width following a
power law with an index of about -2.8. Therefore, it is not surprising
that we find that more variable profiles contain a larger number
of high energy pulses, which are intrinsically narrower and
brighter. \\   

\section{Conclusions}
We have shown there exists a correlation between the
characteristic photon energy in the cosmological rest frame and the
gamma-ray burst variability, as well as the GRB luminosity.
While these
correlations are still tentative, it is reassuring that they are
consistent with several mutually reinforcing trends found between
spectral and temporal properties in a diversity of GRB profiles. 
These relationships can help shed light on the relevant physical mechanisms responsible
for the observed properties of a gamma-ray burst - namely the structure
of the ultra relativistic outflow, the microphysics of shock
acceleration, and the magnetic field generation (see Ramirez-Ruiz \& Lloyd-Ronning 2002).

\begin{acknowledgements}
We would like to thank D. Band, E. Fenimore, D. Lamb and V. Petrosian, and
J. Salmonson for
very useful discussions and comments.
 We also thank the referee for comments that
led to improvements of this paper.  ERR thanks CONACYT, SEP and the ORS for support. 
\end{acknowledgements}

\clearpage

\begin{figure}
\centerline{\psfig{file=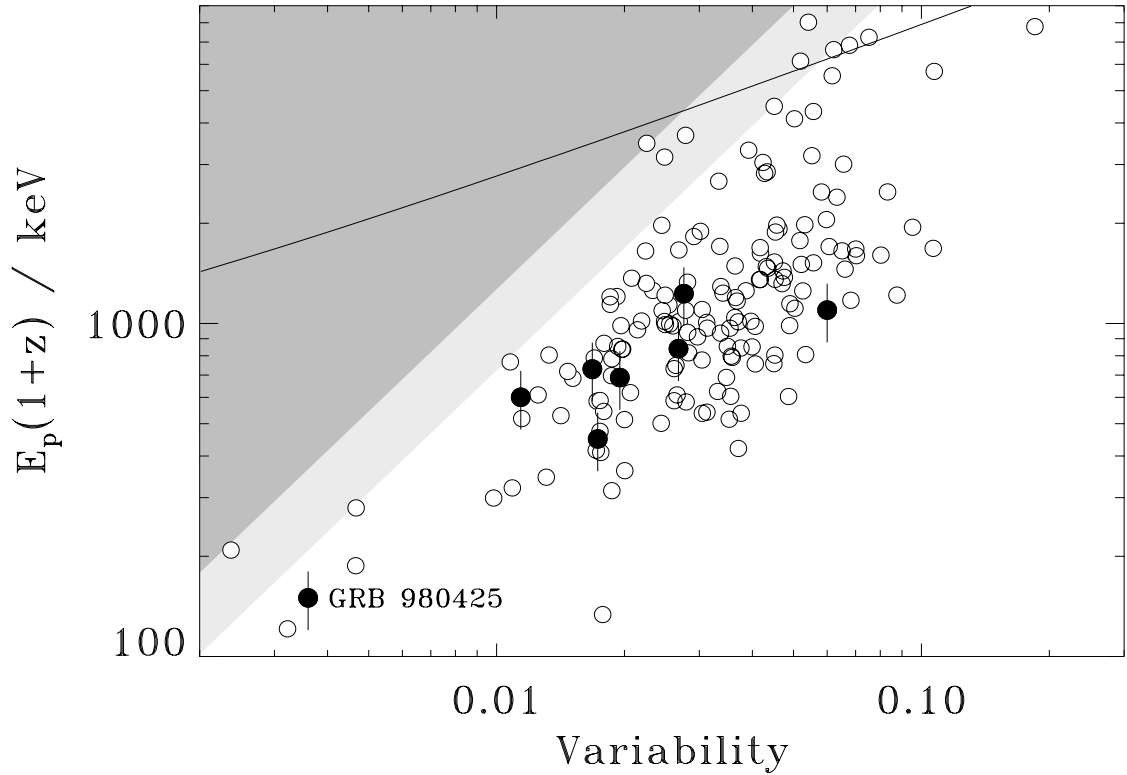,width=0.8\textwidth}}
\caption{The peak of the $\nu F_\nu$ spectrum in the
cosmological rest frame as a function of the burst variability. 
 The filled circles are bursts with secure
redshifts estimates, while the empty circles are bursts in which the
redshift is derived using the variability-luminosity distance
indicator. A trend is clear: the most ``spiky'' bursts  seem to 
have photons with higher characteristic energies.  The shaded regions
and solid line indicate where selection effects may play a role, although 
these effects do not diminish the significance of the correlation.  See \S 3
for details.} 
\label{fig1}
\end{figure}

\clearpage

\begin{figure}
\centerline{\psfig{file=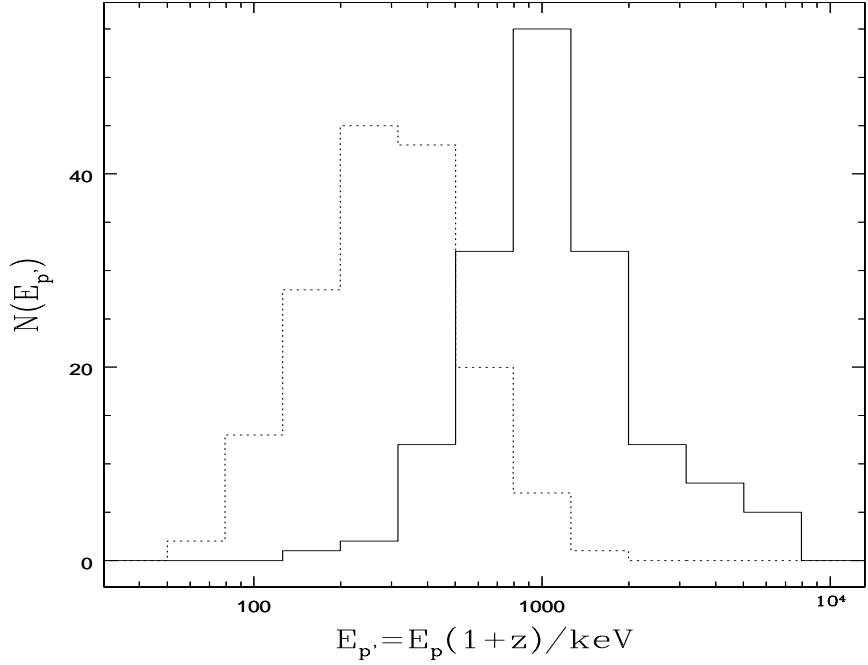,width=0.65\textwidth,height=0.5\textwidth}} 
\caption{Histogram of intrinsic peak energy $E_{p'}=E_{p}(1+z)$
(solid line) for this limited sample. Superposed on the plot (dotted line) is the histogram of
the observed peak energy, $E_{p}$.} 
\label{fig2}
\end{figure}

\begin{figure}
\centerline{\psfig{file=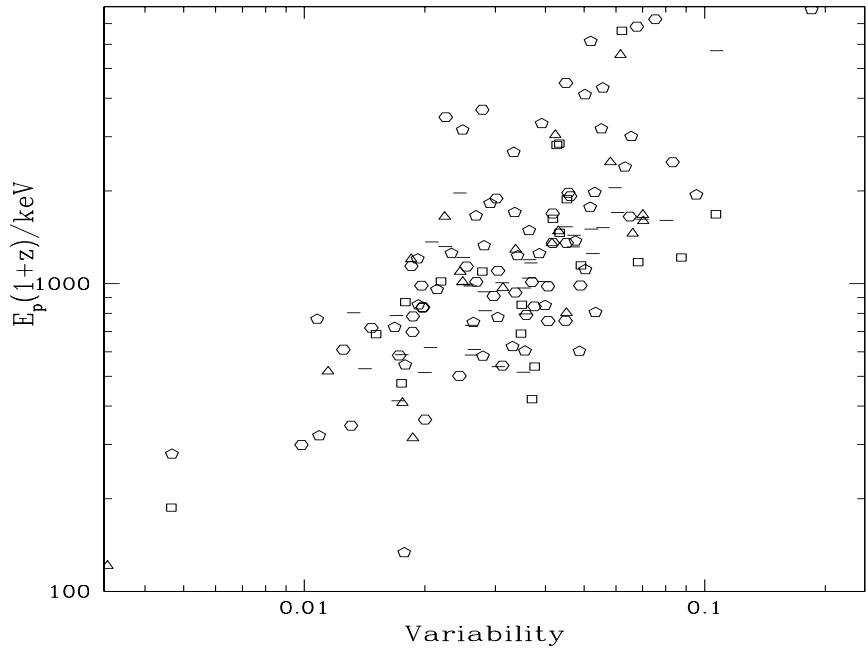,width=0.65 \textwidth,height=0.5\textwidth}} 
\caption{Cosmological rest frame peak energy vs. variability (as in Figure 1), but
dividing the data into flux bins.  The squares, triangles, pentagons, hexagons, and
horizontal lines are
for $f_{p} = 1.5-2.2$ ph/cm$^{2}$/s,  $f_{p} = 2.2-2.7$ ph/cm$^{2}$/s,
 $f_{p}=2.7-4.0$ ph/cm$^{2}$/s, 
$f_{p} = 4.0-7.0$ ph/cm$^{2}$/s, and $f_{p} > 7.0$ ph/cm$^{2}$/s, respectively.} 
\label{fig3}
\end{figure}

\clearpage

\begin{table}  
\begin{center}
\centerline{TABLE 1}
\begin{tabular} {lcccr} \hline \hline
Trigger & $E_{p}^{(1)}$ (keV) & $f_{p}$ (ph/cm$^{2}$/s) & V$^{(2)}$  & z$^{(2)}$ \\ \hline
 109  &  450.30  &   3.62  & 0.02331  &  1.79  \\ 
 130  &  337.10  &   3.47  & 0.01681  &  1.14  \\ 
 143  &  617.70  &  47.57  & 0.05975  &  2.32  \\ 
 219  &  263.40  &  18.06  & 0.01716  &  0.58  \\ 
 249  &  550.90  &  34.62  & 0.01698  &  0.43  \\ 
 394  &  287.90  &   4.78  & 0.01724  &  1.03  \\ 
 398  &  180.40  &   1.71  & 0.01751  &  1.63  \\ 
 467  &  339.30  &   7.73  & 0.03122  &  1.97  \\ 
 503  &  644.10  &   5.05  & 0.07525  & 10.23  \\ 
 563  &  280.10  &   1.89  & 0.01146  &  0.85  \\ 
 660  &  323.00  &   4.55  & 0.03047  &  2.41  \\ 
 676  &  390.70  &   4.20  & 0.02543  &  1.91  \\ 
 678  & 1479.00  &   6.18  & 0.02256  &  1.35  \\ 
 761  &  313.30  &   3.21  & 0.03642  &  3.75  \\ 
 869  &  523.60  &   3.52  & 0.05568  &  7.27  \\ 
 907  &  254.90  &   3.57  & 0.03864  &  3.92  \\ 
 973  &  361.00  &   5.29  & 0.04619  &  4.33  \\ 
1141  &  332.50  &   9.01  & 0.01417  &  0.59  \\ 
1157  &  223.00  &  10.04  & 0.06974  &  6.26  \\ 
1288  &  375.40  &   6.55  & 0.02690  &  1.70  \\ 
1385  &  512.70  &   3.62  & 0.01919  &  1.35  \\ 
1396  &  219.40  &   1.68  & 0.04179  &  6.41  \\ 
1440  &  268.20  &  11.50  & 0.05258  &  3.67  \\ 
1447  &  296.30  &   1.74  & 0.01511  &  1.31  \\ 
1467  &  166.60  &   2.26  & 0.01759  &  1.46  \\ 
1468  &  804.90  &   3.34  & 0.05192  &  6.62  \\ 
1533  &  152.30  &   4.00  & 0.04895  &  5.47  \\ 
1541  &  337.20  &  35.58  & 0.02659  &  0.81  \\ 
1578  &  195.20  &   3.75  & 0.01309  &  0.77  \\ 
1601  &  802.80  &   2.14  & 0.05419  &  9.00  \\ 
1606  &  253.40  &   7.82  & 0.02000  &  1.03  \\ 
1623  &  489.00  &   2.98  & 0.02913  &  2.73  \\ 
1652  &  177.80  &   4.08  & 0.02440  &  1.82  \\ 
1663  &  617.40  &  19.00  & 0.02493  &  0.97  \\ 
1712  &  278.40  &   3.10  & 0.03411  &  3.43  \\ 
1733  &  618.20  &   3.00  & 0.03920  &  4.36  \\ 
1734  &   93.60  &   1.70  & 0.08759  & 12.00  \\ 
1886  &  506.10  &  16.37  & 0.10720  & 10.29  \\ 
1982  &  253.00  &   1.68  & 0.02786  &  3.33  \\ 
1989  &   92.60  &   2.73  & 0.05337  &  7.71  \\ 
1993  &   67.10  &   1.69  & 0.03708  &  5.28  \\ 
2047  &  122.90  &   2.12  & 0.07029  & 12.00  \\ 
2061  &  464.20  &   2.19  & 0.01850  &  1.59  \\
\hline
\end{tabular}
\end{center}
\end{table}
\clearpage
\begin{table}
\begin{center}
\centerline{TABLE 1, {\em continued}}
\begin{tabular} {lcccr} \hline \hline
Trigger & $E_{p}^{(1)}$ (keV) & $f_{p}$ (ph/cm$^{2}$/s) & V$^{(2)}$  & z$^{(2)}$ \\ \hline
2080  &  336.70  &   5.64  & 0.01864  &  1.07  \\ 
2090  &  287.40  &  10.15  & 0.06066  &  4.92  \\ 
2122  &  115.70  &   1.89  & 0.01867  &  1.72  \\ 
2123  &  108.10  &   2.12  & 0.00322  &  0.12  \\ 
2138  &  176.40  &   7.00  & 0.03126  &  2.07  \\ 
2156  &  423.60  &  16.57  & 0.02817  &  1.22  \\ 
2193  &  300.10  &   1.55  & 0.02194  &  2.39  \\ 
2213  &  357.00  &   4.59  & 0.04568  &  4.53  \\ 
2228  &  235.40  &   8.10  & 0.02617  &  1.49  \\ 
2232  &  201.00  &   6.02  & 0.06506  &  7.22  \\ 
2287  &  246.90  &   1.91  & 0.03372  &  4.23  \\ 
2316  &  196.90  &   3.83  & 0.00237  &  0.06  \\ 
2340  &  117.20  &   1.61  & 0.04904  &  8.78  \\ 
2345  &  149.60  &   2.49  & 0.09533  & 12.00  \\ 
2346  &  143.70  &   2.93  & 0.05037  &  6.73  \\ 
2383  &  617.60  &   3.06  & 0.03334  &  3.33  \\ 
2387  &  198.10  &   3.86  & 0.00984  &  0.51  \\ 
2428  &  439.60  &   2.05  & 0.06167  & 11.61  \\ 
2443  &  315.00  &   2.10  & 0.02453  &  2.47  \\ 
2450  &  288.10  &   7.57  & 0.02618  &  1.54  \\ 
2451  &   80.90  &   2.82  & 0.04867  &  6.46  \\ 
2533  &  518.70  &   8.92  & 0.01329  &  0.55  \\ 
2593  &   81.20  &   1.56  & 0.03764  &  5.62  \\ 
2606  &  324.40  &   2.38  & 0.01928  &  1.63  \\ 
2681  &  360.50  &   1.64  & 0.04274  &  6.83  \\ 
2700  &  227.90  &   4.06  & 0.03705  &  3.44  \\ 
2703  &  348.10  &   2.89  & 0.02146  &  1.75  \\ 
2780  &  318.20  &   1.59  & 0.01790  &  1.74  \\ 
2812  &  329.80  &  10.52  & 0.03637  &  2.16  \\ 
2831  &  590.70  &  43.43  & 0.02495  &  0.68  \\ 
2855  &  328.00  &   9.53  & 0.01752  &  0.79  \\ 
2877  &  126.40  &   2.92  & 0.03556  &  3.78  \\ 
2889  &  381.40  &   5.92  & 0.01252  &  0.60  \\ 
2890  &  539.30  &   2.32  & 0.02242  &  2.06  \\ 
2897  &   57.60  &   2.94  & 0.01777  &  1.32  \\ 
2913  &  146.50  &   5.20  & 0.04494  &  4.17  \\ 
2922  &  151.60  &   2.85  & 0.03993  &  4.61  \\ 
2929  &  559.30  &   5.91  & 0.01852  &  1.04  \\ 
2958  &  146.80  &   3.75  & 0.04064  &  4.15  \\ 
2984  &  561.10  &   4.61  & 0.03019  &  2.37  \\ 
2993  & 1029.00  &   3.22  & 0.02486  &  2.07  \\ 
2994  &  956.90  &  14.42  & 0.02448  &  1.06  \\
 3001  &  238.60  &   4.19  & 0.03367  &  2.92  \\ 
 \hline
\end{tabular}
\end{center}
\end{table}
\clearpage
\begin{table}
\begin{center}
\centerline{TABLE 1, {\em continued}}
\begin{tabular} {lcccr} \hline \hline
Trigger & $E_{p}^{(1)}$ (keV) & $f_{p}$ (ph/cm$^{2}$/s) & V$^{(2)}$  & z$^{(2)}$ \\ \hline
3003  &  461.70  &   2.83  & 0.01077  &  0.66  \\ 
3011  &  361.70  &   1.68  & 0.04332  &  6.89  \\ 
3015  &  226.70  &   1.75  & 0.04533  &  7.30  \\ 
3035  &  393.30  &   6.03  & 0.01980  &  1.13  \\ 
3042  &  395.70  &   6.74  & 0.04173  &  3.27  \\ 
3055  &  152.30  &   1.78  & 0.00466  &  0.23  \\ 
3057  &  553.60  &  32.36  & 0.02556  &  0.80  \\ 
3067  &  462.10  &  18.67  & 0.04507  &  2.31  \\ 
3075  &  181.10  &   2.32  & 0.03044  &  3.29  \\ 
3093  &  205.00  &   2.03  & 0.04314  &  6.23  \\ 
3101  &  112.00  &   2.22  & 0.06610  & 12.00  \\ 
3115  &  298.80  &  11.10  & 0.05570  &  4.09  \\ 
3128  &  396.10  &  12.41  & 0.03651  &  2.02  \\ 
3142  &  441.30  &   2.03  & 0.04237  &  5.90  \\ 
3178  &  680.10  &  14.34  & 0.02251  &  0.94  \\ 
3212  &  197.30  &   2.02  & 0.04158  &  5.86  \\ 
3227  &  472.80  &  17.03  & 0.02600  &  1.08  \\ 
3237  &  218.90  &   2.01  & 0.05818  & 10.34  \\ 
3241  &  384.10  &  12.48  & 0.03682  &  2.04  \\ 
3245  &  331.50  &  12.79  & 0.02065  &  0.87  \\ 
3283  &  203.80  &   2.57  & 0.06325  & 10.75  \\ 
3287  &  172.10  &   6.69  & 0.02003  &  1.10  \\ 
3290  &  186.10  &  10.70  & 0.08028  &  7.62  \\ 
3301  &  477.60  &   2.81  & 0.02685  &  2.48  \\ 
3306  &  168.50  &   3.28  & 0.02793  &  2.45  \\ 
3330  &  475.80  &   6.75  & 0.01962  &  1.07  \\ 
3345  &  220.90  &   6.76  & 0.03587  &  2.58  \\ 
3352  &  238.80  &   3.71  & 0.00467  &  0.17  \\ 
3405  &  510.80  &   1.53  & 0.06217  & 12.00  \\ 
3407  &  176.60  &   1.53  & 0.04347  &  7.29  \\ 
3408  &  342.90  &  12.73  & 0.02830  &  1.38  \\ 
3415  &  161.20  &   9.16  & 0.03529  &  2.20  \\ 
3436  &  236.10  &   3.56  & 0.02643  &  2.17  \\ 
3448  &  213.30  &   2.19  & 0.03134  &  3.54  \\ 
3481  &  366.60  &  21.94  & 0.03968  &  1.77  \\ 
3488  &  292.10  &   8.65  & 0.05214  &  4.15  \\ 
3489  &  419.80  &   6.65  & 0.01471  &  0.71  \\ 
3512  &  280.70  &   4.84  & 0.04175  &  3.83  \\ 
3523  &  799.60  &  21.57  & 0.02080  &  0.71  \\ 
3569  &  191.00  &   4.53  & 0.08323  & 12.00  \\ 
3593  &  948.80  &   6.61  & 0.04503  &  3.73  \\ 
3618  &  363.40  &   2.50  & 0.03353  &  3.69  \\ 
3634  &  233.30  &   3.30  & 0.05176  &  6.60  \\ 
 \hline
\end{tabular}
\end{center}
\end{table}
\clearpage
\begin{table}
\begin{center}
\centerline{TABLE 1, {\em continued}}
\begin{tabular} {lcccr} \hline \hline
Trigger & $E_{p}^{(1)}$ (keV) & $f_{p}$ (ph/cm$^{2}$/s) & V$^{(2)}$  & z$^{(2)}$ \\ \hline
3648  &  294.00  &   5.70  & 0.02971  &  2.10  \\ 
3662  &  546.90  &   3.05  & 0.05025  &  6.53  \\ 
3663  &  245.30  &   4.48  & 0.04524  &  4.53  \\ 
3664  &  102.90  &   1.98  & 0.04518  &  6.81  \\ 
3765  &  328.20  &  25.29  & 0.03568  &  1.43  \\ 
3788  &  371.20  &   5.20  & 0.01868  &  1.11  \\ 
3843  &  220.40  &   2.33  & 0.01787  &  1.47  \\ 
3853  &  598.90  &   3.08  & 0.18500  & 12.00  \\ 
3891  &  215.80  &  13.69  & 0.03050  &  1.49  \\ 
3893  &  200.50  &   3.70  & 0.01089  &  0.60  \\ 
3900  &   90.30  &   1.53  & 0.06823  & 12.00  \\ 
3912  &  185.90  &   4.04  & 0.03759  &  3.54  \\ 
3918  &  284.00  &   2.00  & 0.02486  &  2.57  \\ 
3929  &  355.90  &   3.97  & 0.01972  &  1.35  \\ 
3954  &  310.60  &   8.19  & 0.04722  &  3.63  \\ 
4039  & 1249.00  &   5.45  & 0.02785  &  1.94  \\ 
4216  &  114.60  &   1.51  & 0.03475  &  5.01  \\ 
5389  &  197.60  &   4.14  & 0.04062  &  3.96  \\ 
5470  &  707.00  &   4.79  & 0.06769  &  8.68  \\ 
5475  &  328.20  &   2.42  & 0.05522  &  8.72  \\ 
5476  &  135.90  &   2.55  & 0.03313  &  3.60  \\ 
5479  &  188.70  &   2.76  & 0.04757  &  6.29  \\ 
5484  &  246.00  &   2.68  & 0.06557  & 11.23  \\ 
5486  &  303.40  &   9.35  & 0.03542  &  2.19  \\ 
5489  &  301.40  &   9.44  & 0.04699  &  3.37  \\ 
5495  &  128.70  &   2.12  & 0.07006  & 12.00  \\ 
5518  &  213.50  &   2.35  & 0.05314  &  8.27  \\ 
5526  &  386.70  &   3.38  & 0.02814  &  2.44  \\ 
5539  &  129.30  &   1.88  & 0.10660  & 12.00  \\ 
5541  &  147.20  &   1.66  & 0.03494  &  4.80    
\\ \hline \hline
\end{tabular}
\end{center}
\caption{Burst trigger number, observed peak energy, photon peak flux, variability,
and redshift data used in our analysis.}

$^{(1)}$ From fits made to 16 channel data, graciously provided by
Robert S. Mallozzi (deceased).

$^{(2)}$ From Table 2 of Fenimore \& Ramirez-Ruiz (2001).
\end{table}

\begin{table}[t] 
\begin{center}
\centerline{TABLE 2}
\begin{tabular} {lccr} \hline \hline
$L \propto V^{\beta}$ & Eliminate outliers?  & Significance & Function \\ \hline
$\beta=3.3$  &  No & $7\sigma$  & $E_{po}\propto V^{0.8\pm0.15}$  \\
$\beta=3.3$ & Yes & $5.5\sigma$  & $E_{po}\propto V^{0.7\pm0.15}$  \\
$\beta=2.2$  &  No & $5\sigma$  & $E_{po}\propto V^{0.5\pm0.15}$  \\
$\beta=2.2$  &  Yes & $4\sigma$  & $E_{po}\propto V^{0.4\pm0.15}$  \\
$\beta=5.8$  &  No & $8\sigma$  & $E_{po}\propto V^{1.15\pm0.15}$ \\
$\beta=5.8 $ &  Yes & $5.5\sigma$  & $E_{po}\propto V^{1.05\pm0.15}$   
 \\ \hline \hline
\end{tabular}
\end{center}
\caption{Significance and functional form of correlation between
$E_{p'}$ and $V$ given the best fit, upper and lower limit ($\beta = 3.3, \
2.2, \ 5.8$, respectively) for the $L-V$ relation of Fenimore and
Ramirez-Ruiz (2001).  We have computed the
correlation in each case for the two most severe possible truncations (shaded
regions of Figure 1) where the most severe of the two elimates outliers  
(light shaded region in Figure 1).}
\end{table}

\end{document}